\documentclass[12pt]{article}

\newcommand{\be}{\begin{equation}}
\newcommand{\ee}{\end{equation}}
\newcommand{\bi}[1]{\vspace{-3mm} \bibitem{#1}}

\usepackage{epsfig,amsmath,amssymb,graphics,graphicx}

\begin{document}

\begin{center}
International Journal of Theoretical Physics 49 (2010) 293-303
\vskip 5 mm

{\Large \bf Fractional Dynamics of Relativistic Particle}
\vskip 5 mm

{\large \bf Vasily E.Tarasov}\\
\vskip 3mm

{\it Skobeltsyn Institute of Nuclear Physics,\\
Moscow State University, Moscow 119991, Russia} \\
E-mail: tarasov@theory.sinp.msu.ru

\vskip 11 mm

\begin{abstract}
Fractional dynamics of relativistic particle is discussed.
Derivatives of fractional orders with respect to proper time 
describe long-term memory effects that 
correspond to intrinsic dissipative processes. 
Relativistic particle subjected to 
a non-potential four-force is considered as a nonholonomic system.
The nonholonomic constraint in four-dimensional space-time 
represents the relativistic invariance 
by the equation for four-velocity
$u_{\mu} u^{\mu}+c^2=0$, where $c$ is a speed of light in vacuum.
In the general case, the fractional dynamics of relativistic particle 
is described as non-Hamiltonian and dissipative. 
Conditions for fractional relativistic particle 
to be a Hamiltonian system are considered.
\end{abstract}
\end{center}

\vskip 5mm

\noindent
PACS {45.10.Hj;03.30.+p;45.20-d}


\newpage
\section{{Introduction}}

The derivatives of non-integer orders are a natural generalization of 
the ordinary differentiation of integer order. 
Fractional differentiation with respect to time is
characterized by long-term memory effects.
The theory of derivatives and integrals of non-integer order goes back 
to Leibniz, Liouville, Riemann, Grunwald, and Letnikov \cite{SKM,KST}. 
The interest in fractional equations \cite{KST,Podlubny} has been 
growing continually during the last few years because 
of numerous applications in recent studies in mechanics and physics
(for example, see books \cite{Zaslavsky1,CM,SATM} and references therein).

In this paper, we discuss fractional dynamics of relativistic particles 
that are described as nonholonomic systems in four-dimensional space-time.
It is well known that components of the four-velocity
$u^{\mu}=d x^{\mu}/d\tau$ ($\mu=1,2,3,4$ and $\tau$ is a proper time)
are not independent.
The components of the four-velocity are connected
by the equation $u_{\mu} u^{\mu}+c^2=0$, 
where $c$ is a speed of light in vacuum.
This equation allows us to consider the relativistic particle  
as a system with constraint in four-dimensional space-time.
This constraint is nonlinear nonholonomic (nonintegrable) constraint.
As a result, the relativistic invariance for point particles 
is represented by a nonholonomic constraint \cite{KM1,KM2,GM}.

Note that only mechanics of relativistic particles 
can be considered as a mechanics with nonholonomic constraint. 
The relativistic invariance in the field theory 
cannot be represented as a nonholonomic constraint.
At the same time, nonholonomic constraints can be used in the field theory.
For example, the higher spin fields are connected with 
nonholonomic constraints \cite{Sud} and
the gauge fixing conditions for of non-abelian gauge fields
can be described as nonholonomic constraints \cite{NNOO}.
The Euler-Lagrange and Hamilton equations for 
nonholonomic systems in classical field theory are suggested in \cite{K3}.

In the framework of the fractional dynamics, we consider
a relativistic particle subjected to a general four-force. 
In the general case, the four-force is non-potential,
and the  relativistic particle is a non-Hamiltonian system 
in four-dimensional pseudo-Euclidean space-time.
We consider fractional dynamics of non-Hamiltonian and dissipative
systems in relativistic theory.
The fractional equations of motion describes power-law memory effects 
that correspond to intrinsic dissipative processes 
in the relativistic systems.
Note that relativistic particle with dissipation
is discussed in \cite{G1,G2}.
In Refs. \cite{G1,G2}, the Lagrangian and Hamiltonian functions 
for one-dimensional relativistic particles 
with linear dissipation are suggested.
In general, non-Hamiltonian and dissipative $n$-dimensional systems 
with $n > 1$ cannot be described 
by Hamiltonian or Lagrangian since the Helmholtz's conditions 
for these systems are not satisfied \cite{Book}.
In this paper, we consider fractional dynamics of relativistic particles 
as motions of four-dimensional non-Hamiltonian and dissipative systems.


In Sect. 2, the nonholonomic constraint in four-dimensional
space-time for relativistic particle and some notations are considered.
In Sect. 3, we discuss the fractional equations of motion
for relativistic particle that is considered 
as a nonholonomic system.
In Sect. 4, we discuss the d'Alembert-Lagrange principle 
for fractional equations of relativistic particle that is considered 
as a nonholonomic system.
We prove that fractional equations for relativistic systems 
with nonholonomic constraint are represented as 
fractional equations for holonomic systems. 
In Sect. 5, the conditions for fractional relativistic particle 
to be a Hamiltonian or non-dissipative system are considered.
Finally, a short conclusion is given in Sect. 6.

\section{Nonholonomic Constraint}

We consider a four-dimensional pseudo-Euclidean space-time of
points with coordinates $x^{\mu}$:
$x^1=x$, $x^2=y$, $x^3=z$, $x^4=ct$.
The point coordinates in the four-dimensional space-time can be considered
as components {\it radius four-vector} of the point particle,
$\vec{R}=(x^1,x^2,x^3,x^4)=(x,y,z,ct)$.
The square of the elementary radius four-vector
in the four-dimensional space-time is defined by
$(d\vec{R})^2=\eta_{\mu \nu}dx^{\mu}dx^{\nu}$. 
Here and later we mean the sum on the repeated indices
$\mu$ and $\nu$ from 1 to 4.
The coefficients $\eta_{\mu \nu}$ define a metric of pseudo-Euclidean
space-time. This metric is a diagonal tensor such that
$\eta_{11}=\eta_{22}=\eta_{33}=1$ and $\eta_{44}=-1$.
Note that $x_{\mu}$ is not equal to $x^{\mu}$, since
$x_{\mu}=\eta_{\mu \nu} x^{\nu}$ and
$x_1=x^1$, $x_2=x^2$, $x_3=x^3$, and $x_4=-x^4$.

Assume that we have two radius four-vectors $\vec{R}$ and $\vec{R^\prime}$
with coordinates $x^{\mu}$ and $x^{\prime \mu}$ of two reference frames
to describe a relativistic particle.
If the coordinate transformation
$x^{\prime \mu}={a^{\mu}}_{\nu} x^{\nu}$,
where $a^{\mu}_{\nu}$ are constant values, satisfies
the invariant condition:
\be \label{RR1} (d\vec{R}^{\prime})^2=(d\vec{R})^2: \quad
\eta_{\mu \nu}dx^{\prime \mu}dx^{\prime \nu}=
\eta_{\alpha \beta}dx^{\alpha}dx^{\beta},  \ee
then this transformation is a Lorenz transformation.
The coordinates of the radius four-vector
in the proper reference frame are $\vec{R}_{0}=(0,0,0,c\tau)$,
where $\tau$ is a proper time.
Condition (\ref{RR1}) leads us to the relation
\be \label{inv} (d\vec{R})^2=(d\vec{R}_{0})^2: \quad
\eta_{\mu \nu}dx^{\mu} dx^{\nu}=-c^2d\tau^2 . \ee
Using the definition of three-velocity
$v^k=dx^k/dt$, $k=1,2,3$, we get
\be \label{tau} dt=\gamma d\tau, \quad
\gamma=\Bigl(1-v^2/c^2 \bigr)^{-1/2} . \ee

{\it Four-velocity} of the point particle is defined as
a derivative of the radius four-vector with respect to proper time:
\[ \vec{V}=\frac{d\vec{R}}{d\tau}: \quad
u^{\mu}=\frac{dx^{\mu}}{d\tau}. \]
The components of the four-velocity 
$\vec{V}$ are $u^{k}=\frac{dx^{k}}{d\tau}=\gamma v^k$, $k=1,2,3$, and
$u^{4}=\frac{dx^4}{d\tau}=c\gamma$.
Note that rest particles ($\vec{v}=0$) have $u^4=c$.

Equation (\ref{inv}) leads to the relation 
\[ \left(\frac{d\vec{R}}{d\tau}\right)^2=
\left(\frac{d\vec{R}_{0}}{d\tau}\right)^2: \quad
\eta_{\mu \nu}\frac{dx^{\mu}}{d\tau} \frac{dx^{\nu}}{d\tau}=-c^2, \]
which means that square of the four-velocity is a constant value:
$\vec{V}^2=-c^2$.
Therefore we have the constraint equation
\be \label{uu} \eta_{\mu \nu}u^{\mu} u^{\nu}+c^2=0. \ee
As a result, a relativistic particle in the covariant formulation of 
relativistic mechanics is a system with the nonholonomic constraint.
The constraint (\ref{uu}) is nonholonomic since it depends of velocity.
Relativistic mechanics can be considered as nonholonomic mechanics
in the four-dimensional space.


It is known that constraints in mechanics are some simplifications 
of real particle interactions. 
(Note that this statement is not correct in the field theory.)
For example, if we consider the pendulum then we usually
neglect of the forces of thread deformation.
We also neglect of an interaction for constraint (\ref{uu}), 
which defines the relativistic invariance.
If we use the nonholonomic constraint (\ref{uu}),   
then we neglect of a gravity interaction between particles.
Let us consider the deformation of equations (\ref{inv}), (\ref{tau})
and (\ref{uu}) in general theory of relativity \cite{Pauli}.
In the approximation of weak gravity fields, we have
\[ (d\vec{R})^{2}=\eta_{\mu \nu}dx^{\mu} dx^{\nu}-
2\varphi dt^{2}=-c^2d\tau^{2} , \]
where
\[ dt=\gamma^{\prime}d\tau, \quad
\gamma^{\prime}=\Bigl(1+\frac{2\varphi}{c^{2}}
-\frac{v^{2}}{c^{2}}\Bigr)^{-1/2}, \]
and $\varphi$ is a classical (Newtonian) gravity potential.
As a result, we have
\[ \eta_{\mu \nu}u^{\mu} u^{\nu}+c^2 = 2 \varphi \, \gamma^{\prime 2}. \]
Therefore nonholonomic constraint (\ref{uu}), which defines
the relativistic invariance, is connected with the neglect of the gravity
interaction, $\varphi =0$, (in general theory of relativity).

\section{Fractional equations of motion of relativistic particle}

Let $m_0$ be a rest mass of a point relativistic particle.
The {\it four-momentum} of the particle is defined by $\vec{P}=m_{0}\vec{V}$.
The components of the four-momentum are $p^{\mu}=m_{0} u^{\mu}$.
Equation (\ref{uu}) gives
\be \label{4*} \eta_{\mu \nu}p^{\mu} p^{\nu}+m^2_{0}c^2=0 , \ee
In relativistic mechanics the Newtonian equations are replaced by
some generalization, which is invariant under the Lorenz
transformations \cite{Ugarov,Pauli}.
The Newtonian equations are satisfied in the proper reference frame.
The four-vector analog of the Newtonian equations is
\be \label{PF} 
\frac{d \vec{P}}{d\tau}= \vec{{\cal F}} (\tau,\vec{R},\vec{P}). 
\ee
This equation is postulated as a main equation of relativistic dynamics. 
Equation (\ref{PF}) describes a relativistic particle subjected 
to a four-force $ \vec{{\cal F}} = \vec{{\cal F}}(\tau,\vec{R},\vec{P})$. 
Equation (\ref{PF}) must be considered with condition (\ref{4*}).
As a result, we have the equations
\be \label{H1}
\frac{d x^{\mu}}{d \tau}= \frac{1}{m_0} p^{\mu} , \quad
\frac{d p^{\mu}}{d \tau}={\cal F}^{\mu}(\tau,x,p) , \quad
\eta_{\mu \nu}p^{\mu} p^{\nu}+m^2_{0}c^2=0 .
\ee
If $dm_0/d\tau=0$, then (\ref{H1}) give
\be \label{mDx}
m_0 D^2_{\tau} x^{\mu} ={\cal F}^{\mu}(\tau,x,p) .
\ee
where $\eta_{\mu \nu} D^1_{\tau} x^{\mu} D^1_{\tau} x^{\nu}=-c^2$. 
These equations of motion can be generalized for fractional dynamics
to take into account a power-law memory.
We consider a generalization of (\ref{mDx}) in the form of
the fractional differential equations
\be \label{E1}
m_0 \, _0^CD^{\alpha}_{\tau} x^{\mu} = {\cal F}^{\mu} (\tau,x,p),  
\quad (1 < \alpha <2)
\ee
involving the Caputo fractional derivative $ _0^CD^{\alpha}_{\tau}$.
The left-sided Caputo fractional derivative \cite{KST}
of order $\alpha >0$ is defined by
\be \label{Caputo}
\,  _0^CD^{\alpha}_{\tau} x^{\mu}=
\frac{1}{\Gamma(n-\alpha)} \int^{\tau}_0 
\frac{ d\tau' \, D^n_{\tau'} x(\tau')}{(\tau-\tau')^{\alpha-n+1}} =
\, _0I^{n-\alpha}_{\tau} D^n_{\tau} x^{\mu} ,
\ee
where $n-1 < \alpha <n$, $D^n_{\tau}=d^n/d\tau^n$, and $_0I^{\alpha}_{\tau}$ 
is the left-sided Riemann-Liouville fractional integral 
\be \label{FI}
_0I^{\alpha}_{\tau} f( \tau)=\frac{1}{\Gamma(\alpha)} 
\int^{\tau}_0 \frac{f(\tau') d \tau'}{(\tau-\tau')^{1-\alpha}} , \quad (\tau>0).
\ee 
Fractional derivative with respect to proper time describes
a power-law memory effects that 
correspond to intrinsic dissipative processes. 

Using $p^{\mu}=m_0 D^1_{\tau} x^{\mu}$, 
equations (\ref{E1}) can be rewritten in the form 
\be \label{E2}
D^1_{\tau} x^{\mu} =\frac{1}{m_0} p^{\mu} ,
\ee
\be \label{E3}
_0^CD^{\alpha-1}_{\tau} p^{\mu}= {\cal F}^{\mu}(\tau,x,p), \quad (1 <\alpha < 2) ,
\ee
Fractional integration of (\ref{E3}) of order $\alpha -1$ gives 
\be \label{E4}
_0I^{\alpha-1}_{\tau} \, _0^CD^{\alpha-1}_{\tau} p^{\mu} = 
\, _0I^{\alpha-1}_{\tau} {\cal F}^{\mu}(\tau,x,p) .
\ee
Using the fundamental theorem of fractional calculus \cite{FVC}
\[ _0I^{\alpha-1}_{\tau} \, _0^CD^{\alpha-1}_{\tau} p^{\mu}=p^{\mu}(\tau)-p(0) , 
\quad (0 <1-\alpha < 1), \]
we obtain
\be \label{E5}
p^{\mu}(\tau)=p^{\mu}(0)+ \, _0I^{\alpha-1}_{\tau} {\cal F}^{\mu}(\tau,x,p)
\ee
Differentiation of (\ref{E5}) gives
\be \label{E6}
D^1_{\tau} p^{\mu} = \, _0D^{2-\alpha}_{\tau} {\cal F}^{\mu}(\tau,x,p), 
\quad (0<2-\alpha<1) ,
\ee
where $_0D^{2-\alpha}_{\tau}$ is the left-sided Riemann-Liouville fractional derivative 
defined by
\be \label{RLFD}
_0D^{\alpha}_{\tau} x^{\mu}=D^n_{\tau} \ _0I^{n-\alpha}_{\tau} x^{\mu}=
\frac{1}{\Gamma(n-\alpha)} \frac{d^n}{d\tau^n} \int^{\tau}_0 
\frac{x(\tau') d \tau'}{(\tau-\tau')^{\alpha-n+1}} , \quad (n-1 <\alpha \le n) .
\ee

As a result, equation (\ref{E1}) is equivalent to the fractional equations
\be \label{E7}
D^1_{\tau} x^{\mu} = \frac{1}{m_0} p^{\mu} ,
\ee
\be \label{E8}
D^1_{\tau} p^{\mu}= \, _0D^{2-\alpha}_{\tau} {\cal F}^{\mu}(\tau,x,p) , 
\quad (1 <\alpha < 2) .
\ee
These equations describe fractional dynamics of relativistic particle.
Fractional differentiation with respect to proper time is
characterized by long-term memory effects that 
correspond to intrinsic dissipative processes 
in the relativistic systems.

\section{d'Alembert-Lagrange principle for fractional relativistic dynamics}

It is known that the general principle, which allows us to 
derive equations of motion with holonomic and
nonholonomic constraints, is the d'Alembert-Lagrange principle.
For equations (\ref{E7}) and (\ref{E8}) 
this principle leads to the variation equation
\be \label{con1}
\Bigl( \frac{dp^{\mu}}{d\tau} -
\, _0D^{2-\alpha}_{\tau} {\cal F}^{\mu}(\tau,x,p) \Bigr)
\eta_{\mu \nu}\delta x^{\nu}=0. \ee
Multiplying (\ref{E8}) on the variation 
$\delta x_\mu= \eta_{\mu \nu} \delta x^{\nu}$ and summing over $\mu$
we obtain this variational equation.

The variations of coordinates $\delta x^{\mu}$, $\mu=1,...,4$ 
are defined by the relation of the ideal constraint
\be \label{Rk1} 
{\cal R}_{\mu}\delta x^{\mu}=0, 
\ee
where ${\cal R}_{\mu}$ are components of the constraint force vector. 
The four-vector ${\cal R}_{\mu}$ can be considered as a contribution of 
the reaction associated with the constraint to the four-force 
$\, _0D^{2-\alpha}_{\tau} {\cal F}^{\mu}(\tau,x,p)$.
Because a reaction force does no work in a virtual movement 
that is consistent with the corresponding kinematical restriction,
we conclude that ${\cal R}_{\mu}$ must be perpendicular to any $\delta x^{\mu}$ 
that satisfies the constraint equation. 
Thus, if $\delta x^{\mu}$ satisfies constraint equation, 
we have ${\cal R}_{\mu}\delta x^{\mu}=0$.
We now consider which condition $\delta x^{\mu}$ must be realized 
in order to satisfy a constraint equations.
We can derive the usual relativistic equations of motion 
only under the condition (\ref{Rk1}).
For nonholonomic systems a definition of the variations
was suggested by Chetaev \cite{Chet1,Chet2}.
The variations $\delta x^{\mu}$ are defined by the condition:
\be \label{Ch} 
\frac{\partial f}{\partial u^{\mu}} \delta x^{\mu}=0 ,
\ee
where
\be \label{f}
f = \eta_{\mu \nu} u^{\mu} u^{\nu} + c^2 . \ee
Using  (\ref{Rk1}) and (\ref{Ch}), we have  the functions
${\cal R}_{\mu}$ as linear combinations of
$\partial f /\partial u^{\mu}$, i.e.
\[ {\cal R}_{\mu}=\lambda \frac{\partial f}{\partial u^{\mu}}, \]
where $\lambda$ is a Lagrange multiplier.
We note that substitution of (\ref{f}) into (\ref{Ch}) gives
\[ \eta_{\mu \nu} u^{\mu} \delta x^{\nu}=0 . \]

Equations (\ref{con1}) and (\ref{Ch}) give the variational equation
\be \label{Jor2} \left(
\frac{dp^{\mu}}{d\tau} - \, _0D^{2-\alpha}_{\tau} {\cal F}^{\mu}(\tau,x,p)
-\lambda \frac{\partial f}{\partial u^{\mu}} \right)
\delta x^{\mu}=0. \ee
This variational equation is equivalent to 
the fractional equations of motion
\be \label{TT} 
\frac{dp^{\mu}}{d\tau} = \, _0D^{2-\alpha}_{\tau} {\cal F}^{\mu}(\tau,x,p)
+\lambda\frac{\partial f}{\partial u^{\mu}}
\quad (\mu=1,2,3,4) . \ee
We cannot use constraint equation 
for the function $f$ in variational equation 
before the partial derivative on $u^{\mu}$ is taken.

Substitution of (\ref{f}) into (\ref{TT}) gives
the equations of motion 
\be \label{Fuf} 
\frac{d p^{\mu}}{d\tau}=
\, _0D^{2-\alpha}_{\tau} {\cal F}^{\mu}(\tau,x,p) +2\lambda  u^{\mu}, 
\quad u_{\mu}u^{\mu}+c^2=0 , \ee
where $p^{\mu}=m_0u^{\mu}$ and $u^{\mu}=d x^{\mu}/d \tau$.
The system of equations (\ref{Fuf}) is a closed system of 
five equations in the same number of unknowns $x^{\mu}$ and $\lambda$.
Using these equations, we can find the multiplier $\lambda$
as a function $\lambda=\lambda(\tau,x,p)$.
Substituting this function in (\ref{TT}), we get the
equations for coordinates $x^{\mu}$.
It allows us to represent the fractional equations of motion 
for relativistic systems with nonholonomic constraint 
as fractional equations for holonomic systems. 


Differentiating of constraint (\ref{4*}) with respect to $\tau$, we obtain
\be \label{ce2} \eta_{\mu \nu} \frac{dp^{\mu}}{d\tau} p^{\nu}+
m_0\frac{dm_0}{d\tau}c^2=0.\ee
Substituting (\ref{ce2}) in (\ref{Fuf}) 
with $m_0u^{\mu}=p^{\mu}$ and $d m_0/d \tau=0$,  we get
\[ \eta_{\mu \nu} p^{\nu} \, _0D^{2-\alpha}_{\tau} {\cal F}^{\mu}(\tau,x,p) +
2\lambda \eta_{\mu \nu}u^{\mu} p^{\nu}=0. \]
Using the constraint equation
$\eta_{\mu \nu} u^{\mu} u^{\nu}=-c^2$
and the four-momentum $p^{\mu}=m_0 u^{\mu}$, 
we obtain the Lagrange multiplier 
\[ \lambda=\frac{1}{2c^2}(\eta_{\mu \nu} u^{\mu} 
\, _0D^{2-\alpha}_{\tau} {\cal F}^{\nu}(\tau,x,p)) . \]
Therefore the reaction four-force ${\cal R}^{\mu}$ of 
the nonholonomic constraint is 
\[ {\cal R}^{\mu}=2\lambda u^{\mu}=
\frac{1}{c^2} u^{\mu}(u_{\nu} \, _0D^{2-\alpha}_{\tau} {\cal F}^{\nu} ) . \]
As a result, we have the fractional equation
\be \label{FuFu2} 
\frac{dp^{\mu}}{d\tau}= \, _0D^{2-\alpha}_{\tau} {\cal F}^{\mu}(\tau,x,p)
+\frac{1}{c^2} u^{\mu}(u_{\nu} \, _0D^{2-\alpha}_{\tau} {\cal F}^{\nu} ) . 
\ee
These equations define a holonomic system 
subjected to the sum of four-forces 
$\, _0D^{2-\alpha}_{\tau} {\cal F}^{\mu}+{\cal R}^{\mu}$.
If initial dates satisfy constraint equation (\ref{uu}),
then the solution of Eq. (\ref{FuFu2}) describes a fractional dynamics of
the relativistic point particle as a holonomic system.

As a result, we prove the following statement. \\

{\bf Proposition.} 
{\it Fractional equations for the relativistic particle 
subjected to a non-potential four-force ${\cal F}^{\mu}$, which have the form
\be \label{H1nn}
\frac{d x^{\mu}}{d \tau}= \frac{1}{m_0} p^{\mu} , \quad
\frac{d p^{\mu}}{d \tau}=\, _0D^{2-\alpha}_{\tau} {\cal F}^{\mu}(\tau,x,p) , \quad
\eta_{\mu \nu}p^{\mu} p^{\nu}+m^2_{0}c^2=0 .
\ee
with $d m_0 / d \tau=0$, are equivalent to the equations
\be \label{H1n}
\frac{d x^{\mu}}{d \tau}= \frac{1}{m_0} p^{\mu} , \quad
\frac{d p^{\mu}}{d \tau}= \, _0D^{2-\alpha}_{\tau} {\cal F}^{\mu}(\tau,x,p) 
+ {\cal R}^{\mu}(\tau,x,p) ,
\ee
where
\be \label{Rmxp}
{\cal R}^{\mu}(\tau,x,p) = \frac{1}{m^2_0c^2} \, 
p^{\mu} \, (p_{\nu} \, _0D^{2-\alpha}_{\tau} {\cal F}^{\nu}) ,
\ee
and the initial dates satisfy constraint condition (\ref{uu}).} \\

The solution of Eq. (\ref{H1n}) describes the fractional dynamics of
the relativistic particle.

\section{Fractional non-Hamiltonian and dissipative relativistic systems}

The system is called locally Hamiltonian 
if the sum of applied forces satisfies 
the Helmholtz conditions \cite{Helm,TarH}.
If $(x,p) \in {\cal M}$ and ${\cal M}$ is a simply connected region, 
then a locally Hamiltonian system is globally Hamiltonian.
A region is simply connected if it is path-connected
and every path between two points can be continuously
transformed into every other.
A region where any two points can be joined by a path
is called path-connected.

The Helmholtz conditions for fractional equations (\ref{H1n}) have the form
\be \label{HelmCond-2r}
\frac{\partial (\, _0D^{2-\alpha}_{\tau} {\cal F}^{\mu})}{\partial p^{\nu}}+
\frac{\partial {\cal R}^{\mu}}{\partial p_{\nu}}=0,
\ee
\be \label{HelmCond-3r}
\frac{\partial (\, _0D^{2-\alpha}_{\tau} {\cal F}^{\mu})}{\partial x^{\nu}}+
\frac{\partial {\cal R}^{\mu}}{\partial x^{\nu}}-
\frac{\partial (\, _0D^{2-\alpha}_{\tau} {\cal F}^{\nu})}{\partial x^{\mu}}-
\frac{\partial {\cal R}^{\nu}}{\partial x^{\mu}}=0.
\ee
Substitution of (\ref{Rmxp}) into 
Eqs. (\ref{HelmCond-2r}) and (\ref{HelmCond-3r}) gives
\be \label{RHC-1}
\frac{\partial (\, _0D^{2-\alpha}_{\tau} {\cal F}^{\mu})}{\partial p^{\nu}} + 
\frac{1}{m^2_0c^2} 
\frac{\partial [p^{\mu}(p_{\sigma}\, _0D^{2-\alpha}_{\tau} 
{\cal F}^{\sigma})]}{\partial p^{\nu}}=0 ,
\ee
\[
\frac{\partial (\, _0D^{2-\alpha}_{\tau} {\cal F}^{\mu})}{\partial x^{\nu}} +
\frac{1}{m^2_0c^2} p^{\mu} \left(p_{\sigma} 
\frac{\partial (\, _0D^{2-\alpha}_{\tau} {\cal F}^{\sigma})}{\partial x^{\nu}} \right) -
\]
\be \label{RHC-2}
-\frac{\partial (\, _0D^{2-\alpha}_{\tau} {\cal F}^{\nu})}{\partial x^{\mu}}-
\frac{1}{m^2_0c^2} p^{\nu} \left( p_{\sigma} 
\frac{\partial (\, _0D^{2-\alpha}_{\tau} {\cal F}^{\sigma})}{\partial x^{\mu}} \right) 
=0.
\ee
These equations are the Helmholtz conditions \cite{Helm,TarH}
for fractional relativistic dynamics.
If these conditions are satisfied then the 
fractional dynamics of relativistic particle is Hamiltonian.
The fractional relativistic particle subjected 
to a four-force ${\cal F}^{\mu}(\tau,x,p)$ is
non-Hamiltonian if the Helmholtz conditions 
(\ref{RHC-1}) and (\ref{RHC-2}) are not satisfied \cite{Book}.

If 
\[ \Omega(x,p)=\sum^4_{\mu=1} 
\Bigl( \frac{\partial (\, _0D^{2-\alpha}_{\tau} {\cal F}^{\mu})}{\partial p^{\mu}}+
\frac{\partial (\, _0D^{2-\alpha}_{\tau} {\cal R}^{\mu})}{\partial p^{\mu}}
 \Bigr) \ne 0 , \]
then we have a generalized dissipative system \cite{Book}. 
If $\Omega(x,p) \le 0$ for all points $(x,p)$ and
$\Omega(x,p) < 0$ for some points $(x,p)$, 
then the system is a dissipative system.

Note that a one-dimensional relativistic particle with dissipation
is considered in Refs. \cite{G1,G2}.
The Lagrangian and Hamiltonian functions 
for one-dimensional relativistic particles 
with linear dissipation are suggested.
In general, non-Hamiltonian and dissipative $n$-dimensional systems 
with $n > 1$ cannot be described 
by Hamiltonian or Lagrangian since the Helmholtz's conditions 
for these systems are not satisfied \cite{Book}.

In fractional relativistic dynamics 
the principle of stationary action for particle subjected to 
non-potential forces ${\cal F}^{\mu}(\tau,x,p) $ can be used 
if the Helmholtz conditions (\ref{RHC-1}) and (\ref{RHC-2}) are satisfied.
The Hamilton's principle and the principle of stationary action are equivalent 
only for special forms of the four-force ${\cal F}^{\mu}(\tau,x,p)$.
We note that the Hamilton's principle 
is described by nonholonomic variational equation 
\cite{Rum1,RumA,Rum2,Sedov1,Sedov2,Sedov3}.
It allows us to use this principle to obtain 
fractional equations of motion for non-Hamiltonian and dissipative systems. 
The principle of stationary action is defined by 
holonomic variational equation.
Therefore the principle of stationary action cannot be 
to derive fractional equations of motion in the general case.
In general, the Hamilton's principle and nonholonomic variational equations
can be used to describe fractional dynamics of relativistic systems. 
We note that the fractional equations of motion which follow from the 
d'Alembert-Lagrange principle are not equivalent to the fractional equations 
which follow from the principle of stationary action. 
In Refs. \cite{Rum1,RumA,Rum2,Rum3,CR}, authors give proofs 
that the solutions to the equations of motion
which follow from the d'Alembert-Lagrange principle
and the Hamilton's principle do not in general satisfy the equations
which follow from the action principle with nonholonomic constraints. 
The variational Sedov's equation \cite{Sedov1,Sedov2,Sedov3}
(see also \cite{Se1,Chernyi})
can be used in fractional relativistic dynamics 
instead of the principle of stationary action.
We note that relativistic models of continuous media with dissipation
are considered in \cite{Sedov3,Chernyi}. 

Let us consider the four-vector 
$\, _0D^{2-\alpha}_{\tau} {\cal F}^{\mu}$ as the sum 
\be \label{FGP} \, _0D^{2-\alpha}_{\tau} {\cal F}^{\mu}
=G^{\mu}+\Pi^{\mu} , \ee
where $(G^{\mu}u_{\mu})=0$, and $(\Pi^{\mu}u_{\mu})\not=0$.
Substitution of (\ref{FGP}) into (\ref{H1n}) of the form
(\ref{FuFu2}) gives
\[ \frac{dp^{\mu}}{d\tau}=G^{\mu} +\Pi^{\mu}+
\frac{1}{c^2} u^{\mu}(\Pi^{\nu}u_{\nu}). \]
The four-force $G^{\mu}$ is usually called \cite{Ugarov}
a real mechanical force, which satisfies
the orthogonal condition $u_{\mu}G^{\mu}=0$.
The four-vector $\Pi^{\mu}$ describes the energy-momentum
exchange between the point particle and medium.
The components of $\Pi^{\mu}$ are
\[ {\Pi}^{\mu}=(\gamma \vec{\Pi},(\gamma /c) \Phi) , \]
where $\vec{\Pi}$ and $\Phi$ are momentum and energy, 
which are transmitted by convection per unit time.
For the heat transfer, three-momentum $\delta \vec{p}$
and energy $\delta Q$ transmitted per time $d \tau$
are defined by the formulas
$\delta\vec{p}=\vec{\Pi} dt$,  and $\delta Q=\Phi dt$.
The components of $\delta Q^{\mu}$ are
\[ \delta Q^{\mu}=\Pi^{\mu}d\tau=(\delta\vec{p},\frac{1}{c} \delta Q)=
(\gamma \vec{\Pi} \, d\tau, \frac{\gamma}{c}\Phi \, d\tau) , \]
where $\delta Q^{\mu}$ is a four-vector of the heat energy-momentum,
which is transmitted per time $d\tau$.
Note that the value $-u_{\mu}\Pi^{\mu}=-\gamma^2( (\vec{\Pi},\vec{v})-\Phi )$,
is a velocity of the convective transmission
of incoming energy in the rest reference frame.
The four-vectors $G^{\mu}$ and $\Pi^{\mu}$  allow us to describe
non-Hamiltonian and dissipative processes in 
fractional relativistic mechanics.

\section{Conclusion}

We formulate fractional dynamics of relativistic point particle 
as mechanics of the systems with nonholonomic
constraint in the four-dimensional pseudo-Euclidean space-time.
We consider fractional dynamics of 
relativistic particles subjected to four-forces that 
can be non-potential.
The conditions on the four-forces
that allow us to consider fractional dynamics of relativistic particles 
subjected to non-potential forces as Hamiltonian dynamics are suggested.
We prove that the nonholonomic constraint, 
which represents relativistic invariance, 
and the non-potential four-force can be compensated
such that the fractional dynamics is Hamiltonian (and non-dissipative).

Let us note some 
possible extensions of the fractional relativistic dynamics. 

\begin{enumerate}
\item 
Nonholonomic constraints with power-law memory \cite{T4},
which are described by fractional equations,
can be considered in relativistic mechanics 
by using fractional derivatives \cite{KST} with respect to proper time.
\item
The suggested fractional relativistic dynamics  
can be used to generalize quantum theory of non-Hamiltonian 
and dissipative systems \cite{Book}. 
\item
In the framework of the fractional relativistic dynamics 
it is possible to consider a relativistic generalization of the
fractional variational problems \cite{Agrawal1,Agrawal2} 
in Lagrangian and Hamiltonain form \cite{Bal1,Bal2,Bal3,Bal4,Bal5}.
Note that nonholonomic variational equations must be used 
since the fractional equations 
which follow from the d'Alembert-Lagrange principle
(and the Hamilton's principle) do not in general equivalent 
the equations which follow from the action principle 
with nonholonomic constraints \cite{Rum1,RumA,Rum2,Rum3,CR}. 
\end{enumerate}

The study of plasma systems containing ensembles of particles (dust) 
is a rapidly developing field of complex systems research.
One of the general features of complex plasma systems 
is the presence of non-potential interaction forces between 
the dust particles due to the dynamic interaction between 
the dust particles and the plasma 
(for example, see \cite{PP1,PP2,PP3} and references therein). 
In general, these systems cannot be described as Hamiltonian, 
since the energy is not conserved because of the openness of 
the systems due to plasma-particle interaction.
We hope that fractional dynamics of 
relativistic particle subjected to non-potential forces 
can be used to describe relativistic complex plasma systems.


\end{document}